\documentclass[conference]{IEEEtran}
\IEEEoverridecommandlockouts

\usepackage{amsmath,amssymb,amsfonts}
\usepackage{algorithmic}
\usepackage{graphicx}
\usepackage{textcomp}
\usepackage{xcolor}
\usepackage[style=ieee]{biblatex}
\usepackage{setspace}
\usepackage{subcaption}

\begin{document}

\title{Delay Bound Relaxation with Deep Learning-based Haptic Estimation for Tactile Internet}

\author{\IEEEauthorblockN{Georgios Kokkinis$^{*}$, Alexandros Iosifidis$^{\dag}$, Qi Zhang$^{*}$}
\IEEEauthorblockA{\textit{$^{*}$DIGIT and Department of Electrical and Computer Engineering, Aarhus University, Aarhus, Denmark} \\
\textit{$^{\dag}$Faculty of Information Technology and Communication Sciences, Tampere University, Tampere, Finland} \\
Email: \{gkokkinis, qz\}@au.dk; alexandros.iosifidis@tuni.fi
}
}

\maketitle

\begin{abstract}

Haptic teleoperation typically demands sub-millisecond latency and ultra-high reliability (99.999\%) in Tactile Internet. At a 1 kHz haptic signal sampling rate, this translates into an extremely high packet transmission rate, posing significant challenges for timely delivery and introducing substantial complexity and overhead in radio resource allocation. To address this critical challenge, we introduce a novel DL model that estimates force feedback using multi-modal input, i.e. both force measurements from the remote side and local operator motion signals. The DL model can capture complex temporal features of haptic time-series with the use of CNN and LSTM layers, followed by a transformer encoder, and autoregressively produce a highly accurate estimation of the next force values for different teleoperation activities. 
By ensuring that the estimation error is within a predefined threshold, the teleoperation system can safely relax its strict delay requirements. This enables the batching and transmission of multiple haptic packets within a single resource block, improving resource efficiency and facilitating scheduling in resource allocation. Through extensive simulations, we evaluated network performance in terms of reliability and capacity. Results show that, for both dynamic and rigid object interactions, the proposed method increases the number of reliably served users by up to 66\%.

\end{abstract}



\section{Introduction}
\label{sec::Intro}

Bilateral teleoperation with haptic feedback is a cornerstone of the Tactile Internet, which is expected to enable diverse applications such as remote surgery in healthcare, tele-operated driving in transportation, and precision control in Industry 4.0~\cite{FetteweisTI}. In a typical setup, a human operator (leader domain) issues control commands using a high-fidelity haptic device. These commands are transmitted over a network to a remote robot (follower domain), which executes the tasks and sends haptic feedback back to the operator. 

To maintain a stable, immersive interaction, such systems often demand haptic signal sampling rates up to 1 kHz, imposing an end‑to‑end latency constraint on the order of 1 ms with reliability of 99.999\%~\cite{LawrenceKin},~\cite{antonakoglou2018}. 
While ultra‑reliable low‑latency communication (URLLC) in 5G New Radio aims to reduce latency, it is challenging to achieve this over large communication distances or under heavy network load conditions.
It is desirable to have a novel solution that does not rely on such high packet rate of haptic signal, leveraging an edge intelligence engine according to a reference Tactile Internet architecture defined in IEEE P1918.1~\cite{IEEEP1918}.  The edge intelligence engine shall be able to accurately compute the coming haptic packets locally. With this advanced, intelligent function, it will allow significant delay constraint relaxation of haptic packets in Tactile Internet.


Advancements in deep learning have significantly enhanced our ability to model and estimate complex time-series. Recurrent neural networks (RNNs) such as LSTMs and GRUs, have been shown to be able to capture long‑term dependencies in time-series~\cite{Siami2019}, while convolutional networks and attention mechanisms further enhance estimation accuracy by extracting hierarchical features and focusing on salient temporal patterns~\cite{zeng2023}. These advances make it feasible to embed highly accurate, data-driven estimation capabilities into the Edge Intelligence engine, directly within the teleoperation system.

\begin{figure*}[!t]
\label{avg_mse}
 \centering
 \includegraphics[width=0.8\textwidth]{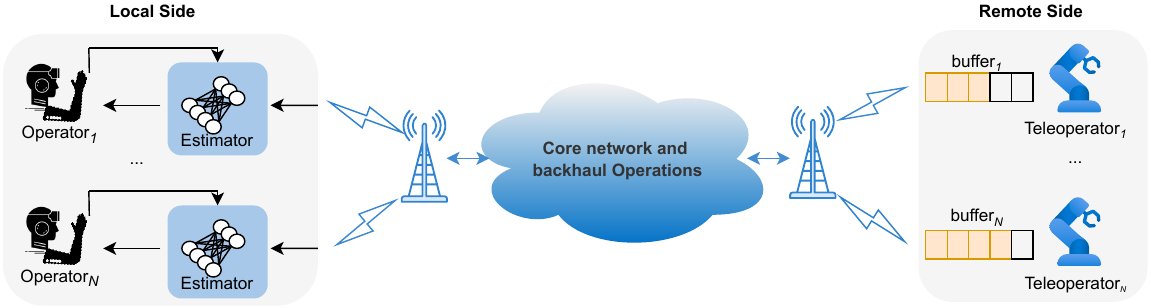}
 \caption{Overall system framework.}
 \label{fig:pipeline}
 \vspace{-1em}
\end{figure*}

Many studies have been conducted to integrate predictive models for teleoperation systems. In the domain of mixed reality, Salvato et al. \cite{Salvato2022} employ a self-attention layer to anticipate hand–object contact from hand-pose patterns, mitigating the actuator latency inherent to post-haptic sensing. While anticipating contact events is relatively straightforward, the estimation continuous force trajectories remains challenging. Xiyan et al. \cite{Xiyan2016} introduce an adaptive estimator with coefficient updates that generates smooth haptic feedback at 1 kHz via frequent sampling and interpolation. However, deterministic methods degrade over longer horizons due to nonlinearity in haptic data. Xu et al. \cite{Xu2020} show that RemedyLSTM, a data-driven recurrent model, outperforms linear estimators in haptic packet estimation and enhances error resilience. Kizilkaya et al. \cite{Kizilkaya2023} evaluate a range of DL-based predictive models in a task-oriented environment augmented with GAN-generated synthetic data. Their approach not only enhances forecasting accuracy but also permits relaxation of bounds of queueing delay in communication systems, enabling more flexible resource allocation schemes. Such advances play an important role in the co-design of control and communication, contributing to ensuring ultra-reliability and low-latency in teleoperation systems.

Recent advances in deep learning enable Transformer‐based models that can leverage both haptic and operator motion data. In this work, we develop a novel deep learning autoregressive model to estimate haptic feedback for the next time window. 
We demonstrate that the latency requirements for haptic packet can be relaxed while maintaining a bounded haptic signal estimation error, thereby enabling more efficient resource allocation, increased network capacity, and improved reliability. 

Our key contributions are summarized as follows.
\begin{itemize}
  \item \textbf{DL Model architecture:} We design and fine‐tune a hybrid DL model incorporating convolutional layers, an LSTM module, and a self‐attention mechanism.
  \item \textbf{Multimodal input integration:} By fusing haptic feedback from the remote robot with the local operator’s command signals (position and velocity), our estimator reduces force estimation error by up to 15\% compared to methods relying solely on past force values.
  \item \textbf{Delay‐bound relaxation and packet batching:} 
With highly accurate force feedback estimation over a defined time horizon, We propose an estimation-and-batching scheme that leverages the relaxed delay bound. This enables multiple haptic packets generated within the relaxed delay bound to be bundled into a single batch and transmitted in a single resource block, significantly simplifying radio resource scheduling and reducing control plane signaling overhead.
  
  \item \textbf{Network capacity improvement:} Extensive experiments were carried out in simulation using geometry-based stochastic channel model. The experimental results demonstrate that with the given radio resource setting, our scheme increases the network capacity by up to 66\%, while ensuring force estimation error bound.
\end{itemize}

For the remainder of the paper, we explain the fundamental relationship between network delay and the stability and transparency of teleoperation, and provide insights into how the Edge Intelligence Engine can help relax these delay constraints in Section \ref{sec::Background}. In Section \ref{sec::multi-modal estimation}, we elaborate our proposed DL model for haptic estimation, along with delay-bound relaxation and packet-batching scheme. Section \ref{sec::Results} presents the performance evaluation of the autoregressive model and simulation results of network performance evaluation, followed by the conclusions in Section \ref{sec::Conclusion}.


\section{Background}
\label{sec::Background}

This section presents background on teleoperation systems and provides insights regarding the effects of delay bound relaxation for communication.

\subsection{Reliability-Stability in Teleoperation}

Bilateral teleoperation involves two parties: the local operator and the remote teleoperator. In a standard setup, the operator sends command signals, typically position and velocity, and receives force and video feedback from the remote environment. As noted in Section~\ref{sec::Intro}, haptic rendering occurs at 1\,kHz, so haptic packets are generated every 1\,ms. This is crucial for stability as the maximum stiffness $k_{\max}$ that can be rendered in a springer-damper system~\cite{Hulin} is formulated by:
\begin{equation}
\label{eq:stiff}
    k_{\max} = \frac{b}{T_s},
\end{equation}
\noindent
where $b$ is the damping of the system, and $T_s$ the sampling period. Another challenge in teleoperation systems is controlling passivity, that is, ensuring that the system does not generate energy~\cite{XiaoXu2016}. Given the formulation of passivity condition in~\cite{XiaoXu2016}, passivity depends directly on the sampling period~$T_s$. 
By increasing~$T_s$, the tolerable round‐trip latency can be relaxed, which has significant meaning in Tactile Internet. In current teleoperation system, latency goes hand in hand with task reliability, in other words, the probability of delay violation directly impacts the task precision and safety. 
Hence, to ensure timely packet delivery, the communication schemes for haptic packet is designed with high reliability requirement to avoid packet retransmission. Now relaxing the latency constraint inherently reduces the corresponding reliability requirements.

While control schemes such as the Wave Variable Transform and Time‐Domain Passivity Approaches (TDPA) extend this relaxation by up to 10\,ms~\cite{Balachandran2016},  without any packet reduction mechanism, the required reliability remains strict, resulting in high packet rate. To mitigate this, Perceptual Deadband techniques based on Weber’s Law~\cite{Steinbach2008} are often employed to suppress imperceptible state changes and reduce transmission frequency:
\begin{equation}
    \frac{\Delta I}{I} = c,
\end{equation}
\noindent
where $\Delta I$ is the minimum perceivable change of the stimulus $I$, and $c$ is the constant Just Noticeable Differences (JND) parameter. Given a JND value of $10\%$, we can reduce the number of haptic packets generated by up to 90\% in some tasks~\cite{Hirche2007}.

\subsection{Queue Delay Bound and Resource Scheduling}

Even in the case of JND-based packet reduction and control schemes such as TDPA, delay bound violations can still occur. As studied in~\cite{Kizilkaya2023}, communication delay consists of a delay in core network and backhaul operations $D_c$, delay due to transmission and propagation $D_t$, and buffer delay in the base station queue $D_q$. In high‐load networks, end‐to‐end delay is dominated by queue delay, where packets queue for resource block allocation. 

Based on a simplified analytical model, where we approximate traffic as a Poisson arrival process, the probability that the queueing delay of a packet $D_{q}$ exceeds a given threshold \(D_{\max}\) can be expressed as

\begin{equation}
    P\bigl(D_{q} > D_{\max}\bigr) =e^{\bigl[-\mu\,(1 - \rho)\,D_{\max}\bigr]},
\end{equation}
where $\rho = \frac{\lambda}{\mu}$ is the system utilization (with \(\lambda\) the average packet arrival rate and \(\mu\) the effective service rate of the channel). This equation reveals a crucial characteristic: the relationship between the delay threshold and the probability of exceeding it is exponential. Specifically, as the acceptable delay threshold $D_{\max}$ is increased, the probability $P(D_{q} > D_{\max})$ decreases exponentially. This implies that slightly relaxing the queueing delay constraint can lead to a significant, non-linear reduction in the likelihood of violating that delay constraint within the framework of this model. This study aims to relax the queue delay threshold by introducing the haptic signal estimator at the local side.



\begin{table}[htbp]
  \caption{Notation}
  \label{tab:notation}
  \centering
  \renewcommand{\arraystretch}{1.15}
  \resizebox{\columnwidth}{!}{%
  \begin{tabular}{ll|ll}
    \hline
    \textbf{Symbol} & \textbf{Meaning} & \textbf{Symbol} & \textbf{Meaning} \\
    \hline
    
    $T_s$ & sampling period & $T$ & input window length \\
    $T_f$ & force input sequence length & $T_c$ & command input sequence length \\
    $t$ & time slot & \text{Linear} & linear transform operation \\
    $\mathcal X_t$ & network input tensor at time $t$ & $\mathcal H_{\mathrm{conv},t}$ & convolutional layer output at time $t$ \\
    $\mathcal H_{\mathrm{lstm},t}$ & LSTM hidden states at time $t$ & $N_{\mathrm{tokens}}$ & number of transformer tokens \\
    $\mathcal\kappa$ & convolution filters & $\mathcal{\omega}$ & convolution filter size \\
    $\mathcal{H}_{tr,t}$ & transformer output at time $t$ &  $d$ & transformer dimension channels \\
    $\hat{\mathbf f}_{t+1}$ & predicted force at $t+1$ & $\mathcal{T}_w$ & relaxed delay constraint \\
    $\mathcal{F}$ & historical force sequence & $\mathcal{C}$ & historical command sequence \\
    $\mathrm{MSE}^{(j)}$ & MSE on the $j$th force axis & $D_c, D_t, D_q$ & core, transmission, queueing delays \\
    $D_{\max}$ & delay bound threshold & $U$ & teleoperation user pairs \\
    $P(D_q > D_{\max})$ & delay violation probability & $S_{RB}$ & size of Resource Block (bytes) \\
    $P$ & packet batch size & $s_p$ & size of haptic packets (bytes) \\
    $\mathcal{K}$ & estimated force sequence length & $\epsilon_{th}$ & maximum error threshold \\
    $\mathcal R$ & packet dropout rate & $N_d$ & number of dropped packets \\
    $I$ & stimulus intensity & $N_g$ & number of generated packets \\
    $\Delta I$ & minimum perceivable change & $c$ & JND constant \\
    $\mathbf f_t$ & force vector at time $t$ & $\mathbf c_t$ & command vector at time $t$ \\
    $n_f$ & number of force features & $n_c$ & number of command features \\
    $b$ & damping coefficient (Ns/m) & $k_{\max}$ & maximum renderable stiffness (N/m) \\
    $n_{\mathrm{lstm}}$ & number of LSTM units & $\epsilon_e$ & teacher forcing probability at epoch $e$ \\
    $E$ & total training epochs & $\lambda$ & average packet arrival rate \\
    $\mu$ & effective service rate & $\rho$ & system utilization \\
    
    \hline
  \end{tabular}%
  }
\vspace{-1em}
\end{table}


\section{Haptic Estimation for Delay Bound Relaxation}
\label{sec::multi-modal estimation}
In this section, we provide a detailed description of the methodology to derive the multi-modal DL-based force estimator, and the new communication scheme for transmitting packets of haptic feedback.

\subsection{Mapping Multi-modal Features to Input Channels}
Beyond relying solely on past force feedback from the remote side, we enhance force estimation by continuously incorporating the operator’s command signals (e.g., position and velocity) at the local side. Each force feedback from the remote robot is, in fact, a response to these transmitted commands, indicating a strong correlation between them. By integrating this multi-modal input, the deep learning model is exposed to more salient and informative patterns, enabling improved estimation accuracy through effective fusion of heterogeneous data streams.

Let the historical force data at time $t$ be represented by a vector $\mathbf{f}_t \in \mathbb{R}^{n_f}$, where $n_f$ is the number of force dimensions. In our experiments, we consider a $3$-dimensional kinesthetic force vector, i.e. $n_f = 3$. We consider a history of $T_f$ time steps, so the historical force input can be represented as a sequence $\mathcal{F} = (\mathbf{f}_{t-T_f+1}, \mathbf{f}_{t-T_f+2}, ..., \mathbf{f}_t)$.

Similarly, the local command data at time $t$ is represented by a vector $\mathbf{c}_t \in \mathbb{R}^{n_c}$, where $n_c$ is the number of command features. The haptic device receives the $3$-dimensional position and $3$-dimensional velocity, i.e. $n_c = 6$. Again, the historical command input can be represented as a sequence $\mathcal{C} = (\mathbf{c}_{t-T_c+1}, \mathbf{c}_{t-T_c+2}, ..., \mathbf{c}_t)$. We choose $T_f = T_c = T$, where $T$ is the input window size; however, $T_f$ and $T_c$ can be different in size.

The input to the neural network at time $t$ is formed by concatenating these historical sequences along the feature dimension. Then this leads to the input tensor at time $t$ be $\mathcal{X}_t \in \mathbb{R}^{T \times (n_f + n_c)}$, having the form:
$$ \mathcal{X}_t = \begin{bmatrix}
\mathbf{f}_{t-T+1}^T & \mathbf{c}_{t-T+1}^T \\
\mathbf{f}_{t-T+2}^T & \mathbf{c}_{t-T+2}^T \\
\vdots & \vdots \\
\mathbf{f}_t^T & \mathbf{c}_t^T
\end{bmatrix}. $$

\begin{figure}[!t]

 \centering
 \includegraphics[width=1\columnwidth]{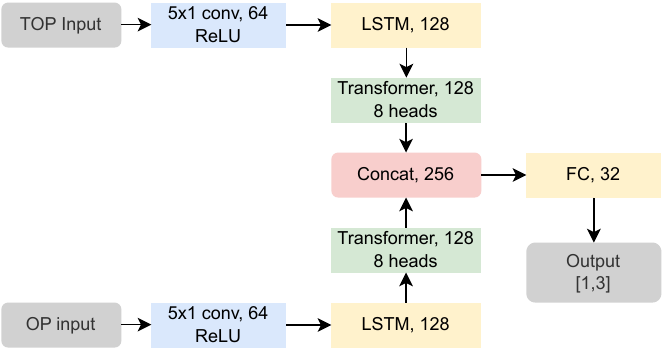}
 \caption{Haptic estimation DL Model architecture.}
 \label{fig:model_fig}
 \vspace{-1em}
\end{figure}

\subsection{DL Model Architecture}

The neural network architecture for haptic force estimation uses a dual‐branch design, as shown in Fig.~\ref{fig:model_fig}. With an input sequence length $T = 100$, the Teleoperator (TOP) branch processes force data, while the Operator (OP) branch handles position and velocity. Each branch independently uses two 1D convolutional layers, with $\kappa=64$ filters and filter size $\omega = 5$, to extract temporal features from input $X_{t}$, yielding $
H_{\mathrm{conv},t} \in \mathbb{R}^{(T-\omega+1)\times \kappa}.
$
These convolutional features then feed into branch‐specific LSTMs, each with $n_{\mathrm{lstm}}=128$ hidden units, producing
$
H_{\mathrm{lstm},t} \in \mathbb{R}^{(T-\omega+1)\times n_{\mathrm{lstm}}}.
$

Each output of the LSTM is treated as one token by the Transformer encoder. Hence:

\begin{equation}
N_{\mathrm{tokens}} = T - \omega + 1,    
\end{equation}

\noindent
with each token embedded in $d = 128$ channels. Using an 8-head self-attention layer, each head attends over those \(N_{\mathrm{tokens}}\) embeddings in parallel, with head dimension
$
d_{\mathrm{head}} = 16.
$

The 128‐dimensional outputs from both branches are concatenated to a 256‐dimensional vector, passed through a 32‐unit fully connected layer, and finally outputting the predicted 3-dimensional force. For single-modal input, i.e. a force‐only mode, the DL model uses only the TOP branch.

\begin{figure}[!t]

 \centering
 \includegraphics[width=1\columnwidth]{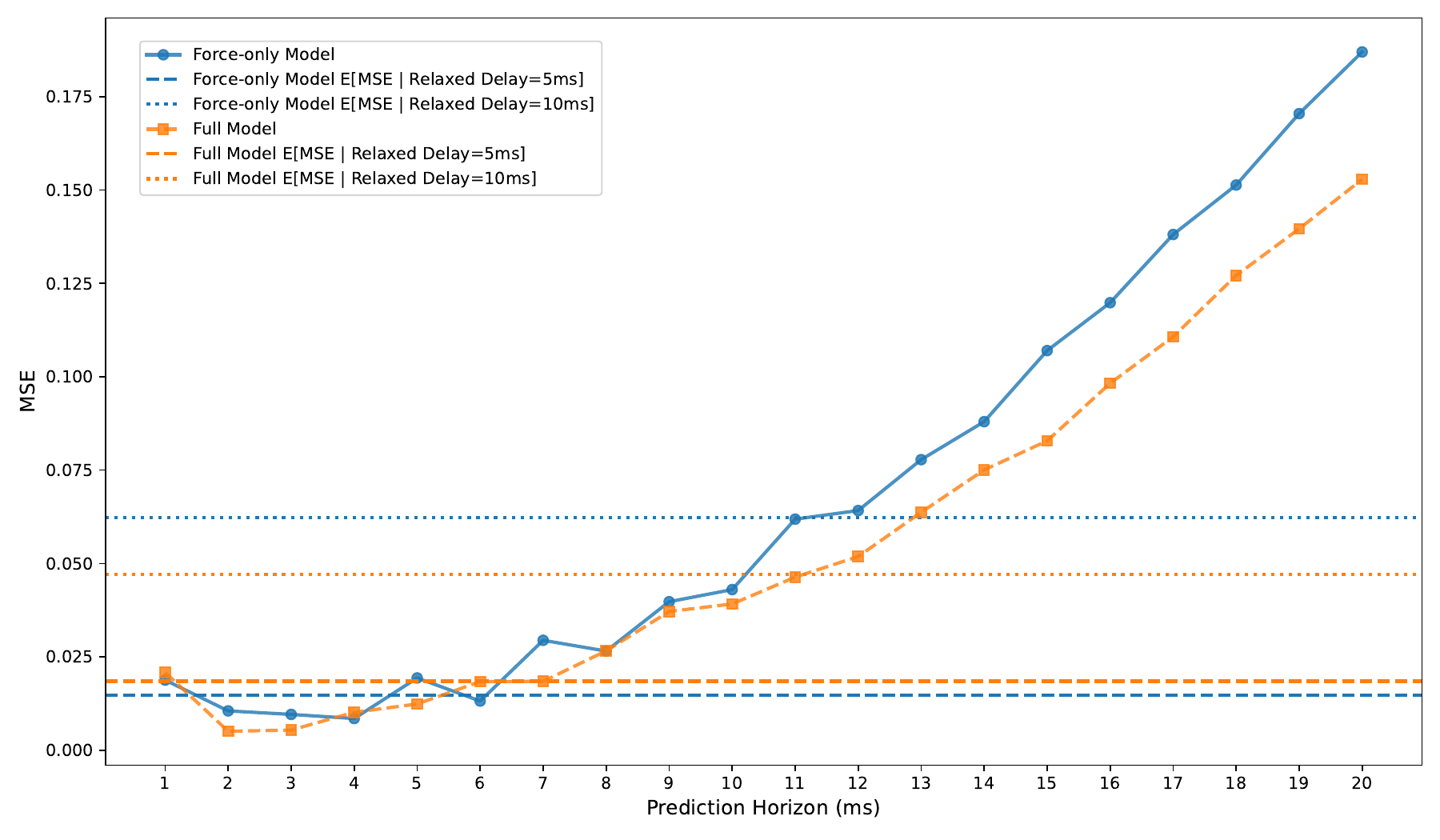}
 \caption{Averaged MSE over all axes.}
 \label{fig:SR_SE}
 \vspace{-1em}
\label{fig:avg_mse}
\end{figure}

\begin{figure}[ht]

 \centering
 \includegraphics[width=1\columnwidth]{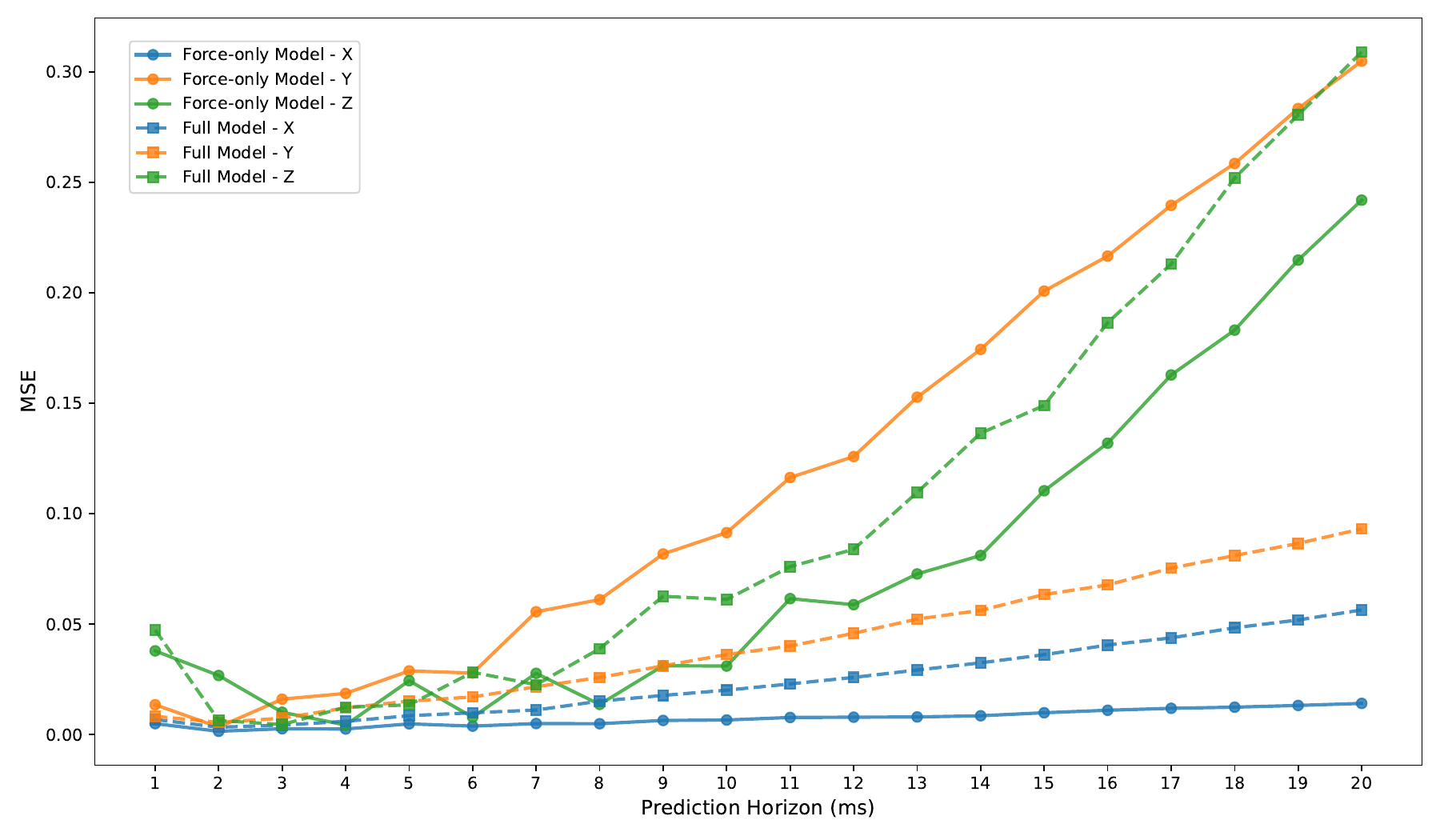}
 \caption{MSE of estimated force per dimension, i.e. (x, y, z).}
 \vspace{-1em}
 \label{fig:per_axis_MSE}
\end{figure}

\subsection{Autoregressive Output}

The predicted force at the next time step is obtained through a linear transformation of the last time step's output from the transformer:
\begin{equation}
\hat{\mathbf{f}}_{t+1} = \text{Linear}(\mathcal{H}_{tr,l,t})    ,
\end{equation}
 where $\mathcal{H}_{tr,l,t}$ the last output token of the transformer.

For the subsequent estimation at time $t+2$, we advance the input window by one time step. The new historical force sequence $\mathcal{F}_{t+1}$ is formed by discarding the oldest force vector $\mathbf{f}_{t-T+1}$, including the predicted force vector $\hat{\mathbf{f}}_{t+1}$, and shifting the remaining true force vectors forward:

\begin{equation}
 \mathcal{F}_{t+1} = (\mathbf{f}_{t-T+2}, ..., \mathbf{f}_t, \hat{\mathbf{f}}_{t+1}).    
\end{equation}

Simultaneously, the historical command sequence is also shifted forward by one time step, incorporating the new true command vector $\mathbf{c}_{t+1}$:

\begin{equation}
\mathcal{C}_{t+1} = (\mathbf{c}_{t-T+2}, ..., \mathbf{c}_t, \mathbf{c}_{t+1}).
\end{equation}

The new input tensor $\mathcal{X}_{t+1}$ for the next estimation step is then formed by concatenating $\mathcal{F}_{t+1}$ and $\mathcal{C}_{t+1}$:

\begin{equation}
\mathcal{X}_{t+1} = \begin{bmatrix}
\mathbf{f}_{t-T+2}^T & \mathbf{c}_{t-T+2}^T \\
\vdots & \vdots \\
\mathbf{f}_t^T & \mathbf{c}_t^T \\
\hat{\mathbf{f}}_{t+1}^T & \mathbf{c}_{t+1}^T
\end{bmatrix}.    
\end{equation}

This sliding window approach, where the predicted force is fed back as input while incorporating the true, up-to-date command information, is crucial for maintaining estimation accuracy over longer estimation horizons. The reliance on true command values at each step helps to mitigate the accumulation of errors that can occur when only predicted force values are used as input.

\begin{figure*}
    \centering
    \begin{subfigure}{0.32\textwidth}
        \centering
        \includegraphics[width=0.95\linewidth]{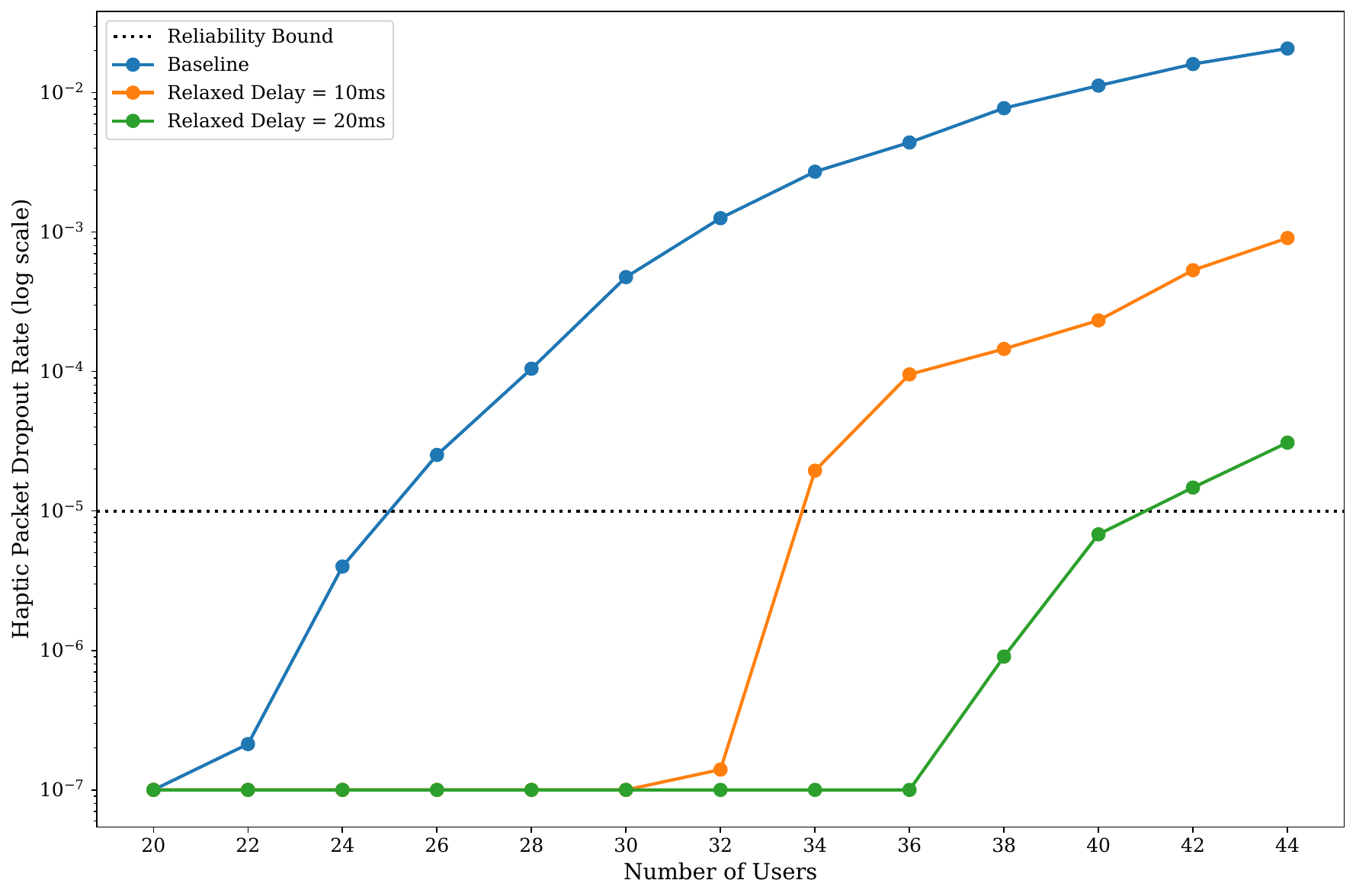}
        \caption{Dynamic Object Pushing}
        \label{fig:sub1}
    \end{subfigure}%
    \hfill
    \begin{subfigure}{0.32\textwidth}
        \centering
        \includegraphics[width=0.95\linewidth]{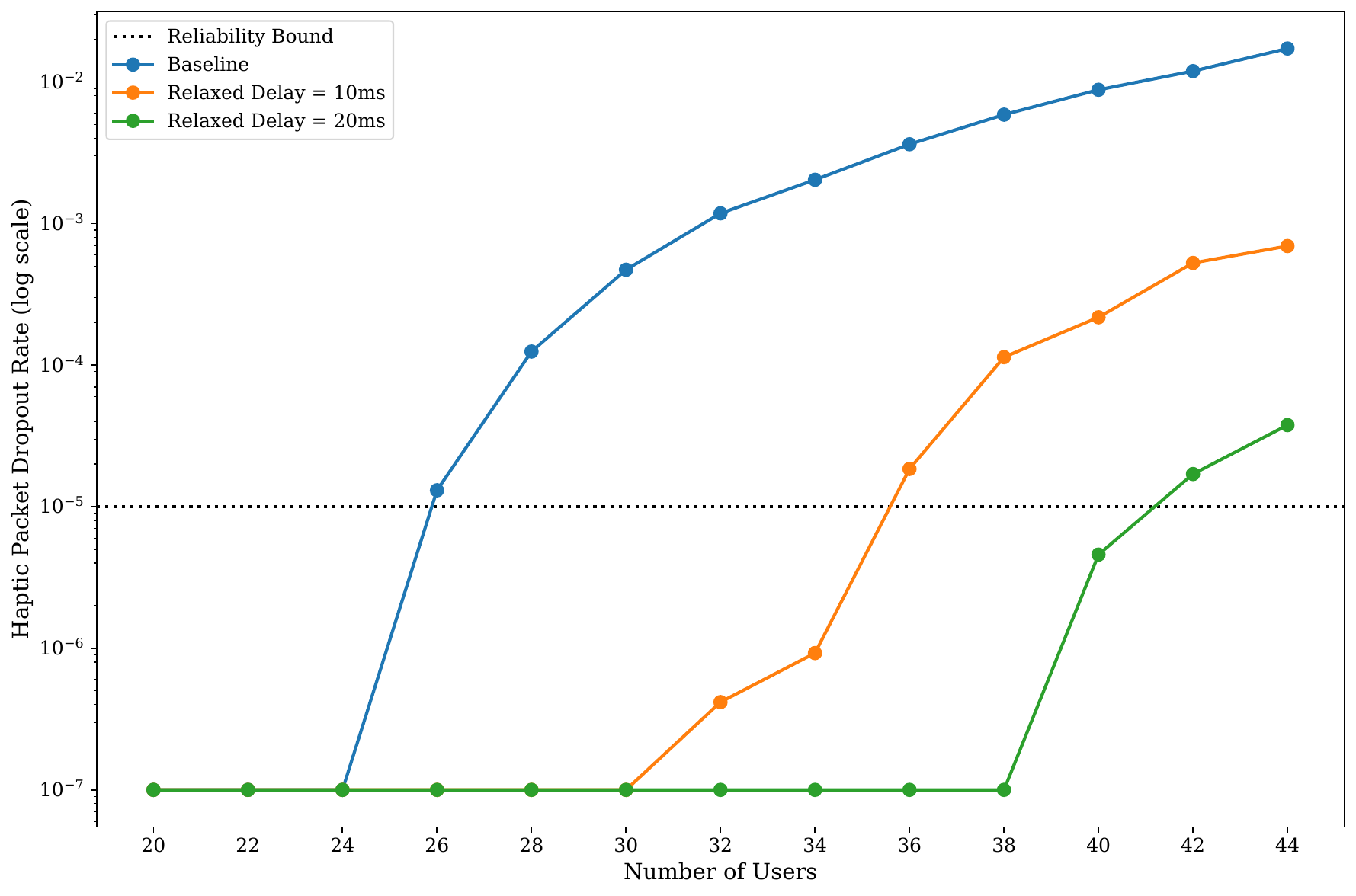}
        \caption{Dynamic Object Tapping}
        \label{fig:sub2}
    \end{subfigure}%
    \hfill
    \begin{subfigure}{0.32\textwidth}
        \centering
        \includegraphics[width=0.95\linewidth]{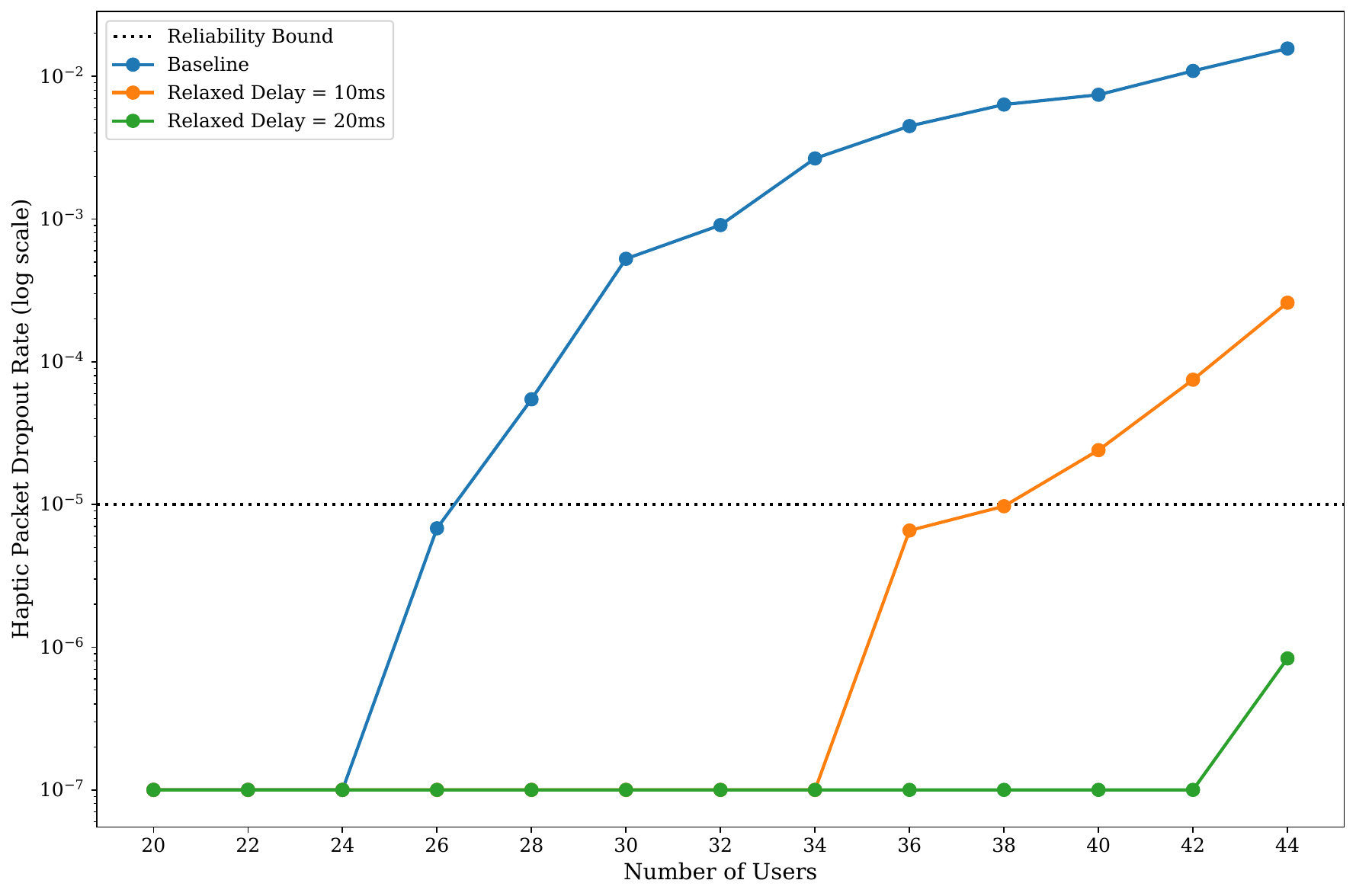}
        \caption{Rigid Body Press and Hold}
        \label{fig:sub3}
    \end{subfigure}
    \caption{Haptic packet dropout rate of different teleoperation activities.}
    \label{fig:dropout_users}
\end{figure*}

\subsection{Delay‐Bound Relaxation and Packet Batching}

In the existing Tactile Internet, each haptic packet must traverse the end-to-end communication system with a hard queuing delay bound $D_{\max}$ to guarantee task stability and reliability. With the autoregressive model, we estimate the next $\mathcal{K}$ force samples $\{\hat{\mathbf f}_{t+1},\dots,\hat{\mathbf f}_{t+\mathcal{K}}\}$ and tolerate an additional delay $\mathcal{T}_w$ such that $\|\hat{\mathbf f}_{t+k} - \mathbf f_{t+k}\|\le \epsilon_{\mathrm{th}}$ for $1\le k\le \mathcal{K}$, where $\epsilon_{\mathrm{th}}$ the predifined maximum error bound. Therefore  $D_{\max} = \mathcal{T}_w$.

Under a relaxed delay bound, multiple generated haptic packets can be packed together and be transmitted in a single resource block (RB) of size \(S_{\mathrm{RB}}\).  If each packet is \(s_p\) bytes, the maximum batch size is
\begin{equation}
  P = \left\lfloor \frac{{\mathcal{T}_w}}{T_s} \right\rfloor
  \,,\quad
  Ps_p \le S_{\mathrm{RB}}. 
\label{eq:Batch}
\end{equation}

By grouping $P$ packets into one RB, it simplifies scheduling in  radio resource allocation and the control‐plane signalling overhead can be significantly reduced, thereby improving spectral efficiency and reducing RB fragmentation. 

After $t = \mathcal{T}_w$, the batched packets are transmitted, delivering the true measurements needed to recalibrate the estimator and ensure the estimation error remains below $\epsilon_{\mathrm{th}}$. This estimation-and-batching scheme thus trades a small, bounded prediction error for a relaxed delay bound and large gains in resource utilization, which is crucial for URLLC services over shared wireless channels.


\section{Experimental Results}
\label{sec::Results}

\subsection{Evaluation of the autoregressive model}

\subsubsection{Training Setup}
We use haptic data traces collected with a Phantom Omni device over 3 fundamental haptic activities \cite{Daniel_traces}: dynamic object pushing, dynamic object tapping, and rigid body press and hold. Each activity is conducted for 120 seconds, with a sampling rate of 1 kHz, yielding 120,000 data points per activity. We use 100,000 steps to avoid artifacts in the beginning and end of each data trace. To validate the performance of the estimator on unseen data, we select the same activities performed twice, one for training and the other for validation. For the evaluation of the estimator, we consider the estimation error only in the validation set.

For training, we use Scheduled Teacher Forcing~\cite{LambTforcing}, a training strategy used in sequence prediction tasks, where the probability of using the true previous force value instead of the model's estimation gradually decreases over training epochs. Let $\epsilon_e$ be the Teacher Forcing probability at epoch $e$. For each estimation step, the input force history uses the true value with probability $\epsilon_e$. A common schedule is:
\begin{equation}
    \epsilon_e = 1 - \frac{e}{E},
\end{equation}

\noindent where $E$ is the total number of epochs. As $e$ increases, the model relies more on its own estimations, bridging the gap to inference.

\subsubsection{Mean Squared Error}

As a performance metric, we use the Mean Squared Error (MSE) along each force dimension, calculated as:
\begin{equation}
\text{MSE}^{(j)} = \left(f^{(j)} - \hat{f}^{(j)}\right)^2,
\end{equation}
\noindent
where $\text{MSE}^{(j)}$ is the Mean Squared Error for force at the $j$-th dimension, $f^{(j)}$ is the true value, and $\hat{f}^{(j)}$ is the estimated value.

The performance of the estimator with respect to MSE is depicted in Fig.~\ref{fig:avg_mse} and Fig.~\ref{fig:per_axis_MSE}, for an estimation horizon of length $\mathcal{T}_w \leq 20\,\text{ms}$. The model with full observation of all modalities performs similarly to using force-only input for small estimation windows, but for $\mathcal{T}_w = 10\,\text{ms}$, the MSE is lower with multi-modal input, with up to $15\%$ reduction in MSE for $\mathcal{T}_w = 20\,\text{ms}$. From Fig.~\ref{fig:per_axis_MSE}, we can derive that while the two models yields similar MSE along the $x$ and $z$ axes, the full model inserts a significantly lower average error on the $y$ axis.

\subsection{Simulation Environment}
The delay violation probability assumes a statistical model for packet arrival rate, however, haptic packet transmission relies on JND, where packet arrival rate cannot be derived from an analytical method. Therefore we conducted the performance evaluation using a simulation, where each User Equipment (UE) generates haptic packets based on the 3 teleoperation activities, with a JND parameter of 10\%. We simulate $U$ teleoperation user pairs, each comprising an operator that generates motion signal packets and a teleoperator that generates both haptic and video traffic. For video traffic generation, we consider a statistical model according to standard requirements for teleoperation~\cite{Stotko2019},~\cite{AV1_2018}. We assume a bandwidth $B$ of $10$ MHz split into 100 RBs, with 90\% allocated for video and 10\% for haptic data, since video has higher bandwidth requirements. The users are served every 1 ms with round robin scheduling. For channel modeling, we utilize the channel profiles from~\cite{Intent2024}. The profiles were generated using Quadriga, a stochastic channel simulator, which includes experimentally validated channel models.
\subsection{Performance analysis}

For the first round of experiments, we used the average packet dropout rate $\mathcal{R}$ as a performance metric:


\begin{equation}
    \mathcal{R} = \frac{1}{U}\sum_{u=0}^{U}\frac{N^{u}_{\text{d}}} {N^{u}_{\text{g}}} ,
\label{eq:dropout}
\end{equation}
\noindent
where  $N^{u}_{\text{d}}$  and $N^{u}_{\text{g}}$ are the number of haptic packets dropped and generated per user, respectively. The packets are dropped if the amount of time that they remain in the buffer is greater than the delay bound. 

In Fig.~\ref{fig:dropout_users}, we present the dropout rate for varying number of users across three teleoperation activities. As baseline, we assume that the packet is dropped if it is not transmitted at the first time step. For dynamic object pushing, up to 24 users can be served with the reliability $99.999\%$, without delay relaxation. With haptic estimation, the network capacity is increased to 32 and 40 users, respectively, when allowing 10~ms and 20~ms delay relaxation. This is equivalent to 33\% and 66\% capacity increase. A similar trend is observed for dynamic object tapping. For rigid body interaction, which involves fewer environmental changes, slightly more users can be supported.

\begin{figure}[t!]

 \centering
 \includegraphics[width=1\columnwidth]{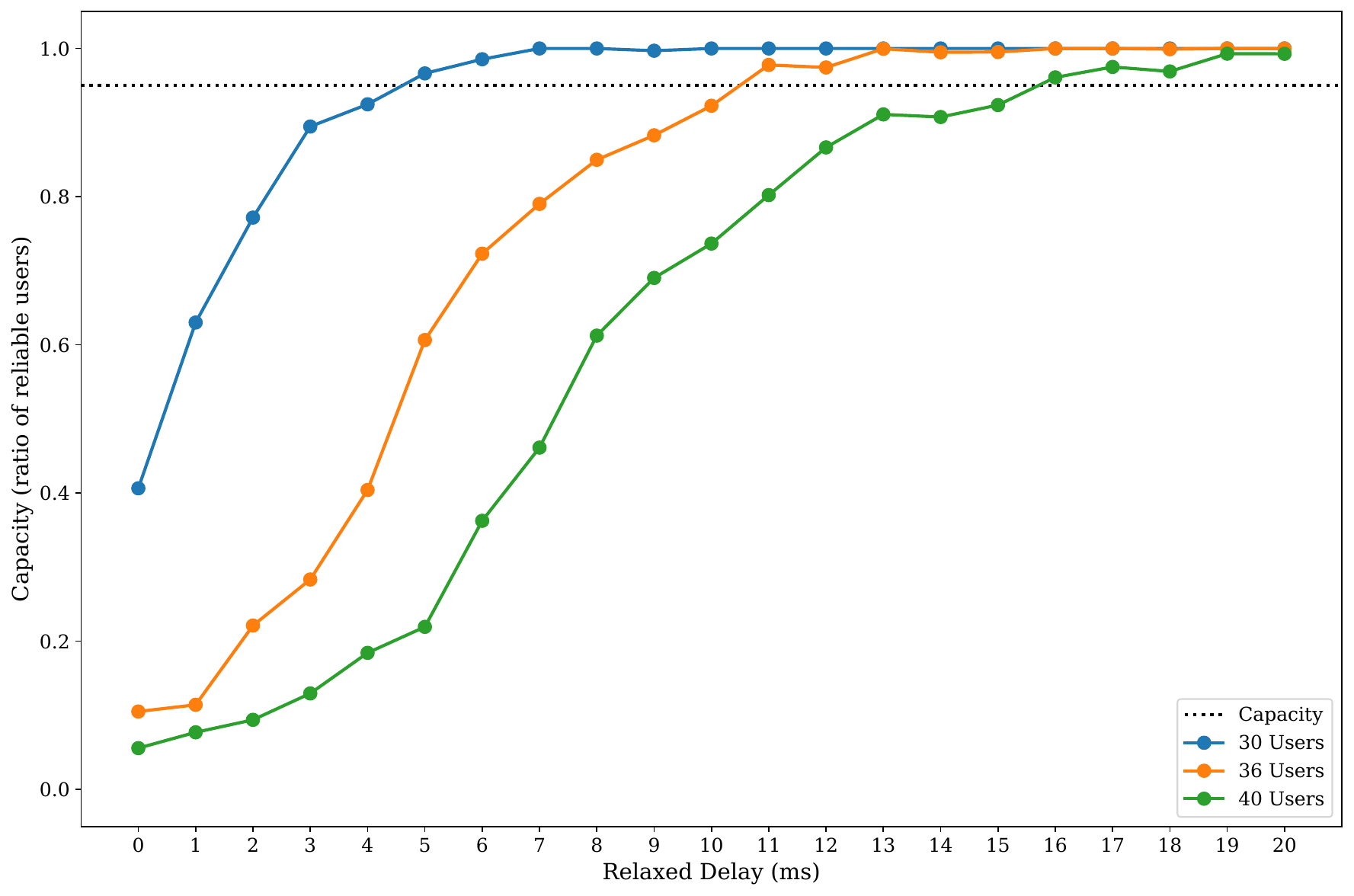}
 \caption{Capacity over delay bound relaxation.}
 \vspace{-1em}
 \label{fig:capacity}
\end{figure}

We define network capacity as the number of users in the network, given that $95\%$ of the users are statisfied. In Fig.~\ref{fig:capacity}, the change of network capacity as a function of varying delay bound relaxation is presented for the scenario of dynamic object pushing. For $\mathcal T_w = 5\,\text{ms}$, the network can accommodate 30 users. For $\mathcal T_w = 10\,\text{ms}$ and $\mathcal T_w = 15\,\text{ms}$, the capacity is increased to 36 and 40 users, respectively. Overall, by relaxing the delay bound from $\mathcal T_w = 5\,\text{ms}$ to $\mathcal T_w = 15\,\text{ms}$, we achieve a $25\%$ increase in number of satisfied users.

\section{Conclusion}
\label{sec::Conclusion}

In this paper, we develop a novel DL model for haptic signal estimation with multi-modal input, which achieves very low MSE of estimated force feedback. With this breakthrough, it is feasible to allow the teleoperation system to relax strict delay constraint for haptic packets. By relaxing the delay bound, a sequence of haptic packets can be bundled together and be transmitted in a single resource block. This facilitates better scheduling in radio resource allocation, and improves resource utilization, thereby increasing network capacity while ensuring QoS and task reliability. As future work, we aim to incorporate this edge intelligence engine into our Tactile Internet testbed.

\section*{Acknowledgment}

This research was supported by the TOAST project, funded by the European Union’s Horizon Europe research and innovation program under the Marie Skłodowska-Curie Actions Doctoral Network (Grant Agreement No. 101073465), the Danish Council for Independent Research project eTouch (Grant No. 1127- 00339B) and NordForsk Nordic University Cooperation on Edge Intelligence (Grant No. 168043).

\begingroup
\renewcommand*{\bibfont}{\small}
\setlength{\bibitemsep}{1pt}  
\setstretch{0.85}  
\printbibliography

@ARTICLE{IEEEP1918,
  author={Holland, Oliver and Steinbach, Eckehard and Prasad, R. Venkatesha and Liu, Qian and Dawy, Zaher and Aijaz, Adnan and Pappas, Nikolaos and Chandra, Kishor and Rao, Vijay S. and Oteafy, Sharief and Eid, Mohamad and Luden, Mark and Bhardwaj, Amit and Liu, Xun and Sachs, Joachim and Araújo, José},
  journal={Proceedings of the IEEE}, 
  title={The IEEE 1918.1 “Tactile Internet” Standards Working Group and its Standards}, 
  year={2019},
  volume={107},
  number={2},
  pages={256-279},
  doi={10.1109/JPROC.2018.2885541}}

@article{Intent2024,
  author={Nahum, Cleverson Veloso and Lopes, Victor Hugo L. and Dreifuerst, Ryan M. and Batista, Pedro and Correa, Ilan and Cardoso, Kleber Vieira and Klautau, Aldebaro and Heath, Robert W.},
  journal={IEEE Transactions on Wireless Communications}, 
  title={Intent-Aware Radio Resource Scheduling in a RAN Slicing Scenario Using Reinforcement Learning}, 
  year={2024},
  volume={23},
  number={3},
  pages={2253-2267}}

@article{antonakoglou2018,
  author={Antonakoglou, Konstantinos and Xu, Xiao and Steinbach, Eckehard and Mahmoodi, Toktam and Dohler, Mischa},
  journal={IEEE Communications Surveys \& Tutorials}, 
  title={Toward Haptic Communications Over the 5G Tactile Internet}, 
  year={2018},
  volume={20},
  number={4},
  pages={3034-3059}}

@article{XiaoXu2016,
  author={Xu, Xiao and Schuwerk, Clemens and Cizmeci, Burak and Steinbach, Eckehard},
  journal={IEEE Transactions on Haptics}, 
  title={Energy Prediction for Teleoperation Systems That Combine the Time Domain Passivity Approach with Perceptual Deadband-Based Haptic Data Reduction}, 
  year={2016},
  volume={9},
  number={4},
  pages={560-573}}

@INPROCEEDINGS{AV1_2018,
  author={Chen, Yue and Murherjee, Debargha and Han, Jingning and Grange, Adrian and Xu, Yaowu and Liu, Zoe and Parker, Sarah and Chen, Cheng and Su, Hui and Joshi, Urvang and Chiang, Ching-Han and Wang, Yunqing and Wilkins, Paul and Bankoski, Jim and Trudeau, Luc and Egge, Nathan and Valin, Jean-Marc and Davies, Thomas and Midtskogen, Steinar and Norkin, Andrey and de Rivaz, Peter},
  booktitle={2018 Picture Coding Symposium (PCS)}, 
  title={An Overview of Core Coding Tools in the AV1 Video Codec}, 
  year={2018},
  volume={},
  number={},
  pages={41-45}}

@INPROCEEDINGS{Stotko2019,
  author={Stotko, Patrick and Krumpen, Stefan and Schwarz, Max and Lenz, Christian and Behnke, Sven and Klein, Reinhard and Weinmann, Michael},
  booktitle={2019 IEEE/RSJ International Conference on Intelligent Robots and Systems (IROS)}, 
  title={A VR System for Immersive Teleoperation and Live Exploration with a Mobile Robot}, 
  year={2019},
  volume={},
  number={},
  pages={3630-3637}}

@article{Hirche2007,
  author={Hirche, Sandra and Hinterseer, Peter and Steinbach, Eckehard and Buss, Martin},
  journal={Presence}, 
  title={Transparent Data Reduction in Networked Telepresence and Teleaction Systems. Part I: Communication without Time Delay}, 
  year={2007},
  volume={16},
  number={5},
  pages={523-531}}

@misc{Daniel_traces,
  author    = {Daniel Rodriguez-Guevara},
  title     = {Kinesthetic Data Traces},
  howpublished = {Dataset},
  note = {Avaliable: https://cloud.lkn.ei.tum.de/s/M7xWrCecdYYZJsw}
}

@ARTICLE{FetteweisTI,
  author={Fettweis, Gerhard P.},
  journal={IEEE Vehicular Technology Magazine}, 
  title={The Tactile Internet: Applications and Challenges}, 
  year={2014},
  volume={9},
  number={1},
  pages={64-70},
  doi={10.1109/MVT.2013.2295069}}

@ARTICLE{LawrenceKin,
  author={Lawrence, D.A.},
  journal={IEEE Transactions on Robotics and Automation}, 
  title={Stability and transparency in bilateral teleoperation}, 
  year={1993},
  volume={9},
  number={5},
  pages={624-637},
  doi={10.1109/70.258054}}

@ARTICLE{Salvato2022,
  author={Salvato, M. and Heravi, Negin and Okamura, Allison M. and Bohg, Jeannette},
  journal={IEEE Robotics and Automation Letters}, 
  title={Predicting Hand-Object Interaction for Improved Haptic Feedback in Mixed Reality}, 
  year={2022},
  volume={7},
  number={2},
  pages={3851-3857},
  doi={10.1109/LRA.2022.3148458}}

@article{Xiyan2016,
  author       = {Xiyuan Hou and
                  Olga Sourina},
  title        = {Real-time Adaptive Prediction Method for Smooth Haptic Rendering},
  journal      = {CoRR},
  volume       = {abs/1603.06674},
  year         = {2016},
  url          = {http://arxiv.org/abs/1603.06674},
  eprinttype    = {arXiv},
  eprint       = {1603.06674},
  bibsource    = {dblp computer science bibliography, https://dblp.org}
}

@INPROCEEDINGS{Xu2020,
  author={Xu, Yiwen and Zheng, Quanfei and Lin, Qingxu and Wang, Kai and Zhao, Tiesong},
  booktitle={2020 IEEE 6th International Conference on Computer and Communications (ICCC)}, 
  title={Error Resilience Algorithm for Haptic Communication Based on Remedy-LSTM}, 
  year={2020},
  volume={},
  number={},
  pages={2207-2211},
  doi={10.1109/ICCC51575.2020.9345145}}

@ARTICLE{Kizilkaya2023,
  author={Kizilkaya, Burak and She, Changyang and Zhao, Guodong and Imran, Muhammad Ali},
  journal={IEEE Transactions on Vehicular Technology}, 
  title={Task-Oriented Prediction and Communication Co-Design for Haptic Communications}, 
  year={2023},
  volume={72},
  number={7},
  pages={8987-9001},
  doi={10.1109/TVT.2023.3247442}}

@inproceedings{LambTforcing,
 author = {Lamb, Alex M and ALIAS PARTH GOYAL, Anirudh Goyal and Zhang, Ying and Zhang, Saizheng and Courville, Aaron C and Bengio, Yoshua},
 booktitle = {Advances in Neural Information Processing Systems},
 editor = {D. Lee and M. Sugiyama and U. Luxburg and I. Guyon and R. Garnett},
 pages = {},
 publisher = {Curran Associates, Inc.},
 title = {Professor Forcing: A New Algorithm for Training Recurrent Networks},
 url = {https://proceedings.neurips.cc/paper_files/paper/2016/file/16026d60ff9b54410b3435b403afd226-Paper.pdf},
 volume = {29},
 year = {2016}
}

@INPROCEEDINGS{Hulin,
  author={Hulin, Thomas and Preusche, Carsten and Hirzinger, Gerd},
  booktitle={2006 IEEE/RSJ International Conference on Intelligent Robots and Systems}, 
  title={Stability Boundary for Haptic Rendering: Influence of Physical Damping}, 
  year={2006},
  volume={},
  number={},
  pages={1570-1575},
  doi={10.1109/IROS.2006.282043}}

@ARTICLE{Steinbach2008,
  author={Hinterseer, Peter and Hirche, Sandra and Chaudhuri, Subhasis and Steinbach, Eckehard and Buss, Martin},
  journal={IEEE Transactions on Signal Processing}, 
  title={Perception-Based Data Reduction and Transmission of Haptic Data in Telepresence and Teleaction Systems}, 
  year={2008},
  volume={56},
  number={2},
  pages={588-597},
  doi={10.1109/TSP.2007.906746}}

@INPROCEEDINGS{Balachandran2016,
  author={Balachandran, Ribin and Artigas, Jordi and Mehmood, Usman and Ryu, Jee-Hwan},
  booktitle={2016 IEEE/RSJ International Conference on Intelligent Robots and Systems (IROS)}, 
  title={Performance comparison of Wave Variable Transformation and Time Domain Passivity Approaches for time-delayed teleoperation: Preliminary results}, 
  year={2016},
  volume={},
  number={},
  pages={410-417},
  doi={10.1109/IROS.2016.7759087}}

@misc{zeng2023,
      title={Financial Time Series Forecasting using CNN and Transformer}, 
      author={Zhen Zeng and Rachneet Kaur and Suchetha Siddagangappa and Saba Rahimi and Tucker Balch and Manuela Veloso},
      year={2023},
      eprint={2304.04912},
      archivePrefix={arXiv},
      primaryClass={cs.LG},
      url={https://arxiv.org/abs/2304.04912}, 
}

@INPROCEEDINGS{Siami2019,
  author={Siami-Namini, Sima and Tavakoli, Neda and Namin, Akbar Siami},
  booktitle={2019 IEEE International Conference on Big Data (Big Data)}, 
  title={The Performance of LSTM and BiLSTM in Forecasting Time Series}, 
  year={2019},
  volume={},
  number={},
  pages={3285-3292},
  doi={10.1109/BigData47090.2019.9005997}}
\endgroup
\end{document}